\documentclass[conference]{IEEEtran}
\IEEEoverridecommandlockouts
\usepackage{cite}
\usepackage{amsmath,amssymb,amsfonts}
\usepackage{graphicx}
\usepackage{textcomp}
\usepackage{xcolor}
\usepackage{algorithm}
\usepackage[noend]{algpseudocode}
\usepackage{makecell}
\usepackage{array}
\usepackage{url}
\usepackage{subcaption}
\usepackage{pifont}
\newcommand{\cmark}{\ding{51}}
\newcommand{\xmark}{\ding{55}}
\usepackage{tablefootnote}
\usepackage{threeparttable}
\usepackage{enumitem}

\def\BibTeX{{\rm B\kern-.05em{\sc i\kern-.025em b}\kern-.08em
    T\kern-.1667em\lower.7ex\hbox{E}\kern-.125emX}}
\begin{document}

\title{V-TSN: A Software-Defined TSN Overlay for General-Purpose Networks}

\author{
\IEEEauthorblockN{
Mohammadparsa Karimi$^{1}$,
Majid Nabi$^{1}$,
Ahmed Khalaf$^{2}$,
Andrew Nelson$^{1}$,
Kees Goossens$^{1}$,
Twan Basten$^{1}$
}

\IEEEauthorblockA{$^{1}$Department of Electrical Engineering, Eindhoven University of Technology, The Netherlands\\
Emails: \{m.karimi, m.nabi, a.t.nelson, k.g.w.goossens, a.a.basten\}@tue.nl}

\IEEEauthorblockA{$^{2}$Architecture and Network Solutions, AUMOVIO, Germany\\
Email: ahmed.khalaf@aumovio.com}
}

\maketitle

\begin{abstract}
Time-Sensitive Networking (TSN) extends Ethernet with deterministic communication for time-critical applications such as industrial automation, in-vehicle networks, and cyber-physical systems. However, realizing TSN behavior without dedicated hardware is difficult. During design and validation, offline simulation cannot run application software at real-time speed when costly specialized TSN hardware is not (yet) available. At deployment time, many systems run on general-purpose and cloud networks with no native TSN support, where provisioning full TSN hardware is unnecessary or impractical for applications that tolerate relaxed timing. In this paper, we introduce Virtual Time-Sensitive Networking (V-TSN), a software-defined overlay that realizes gPTP-based synchronization and TSN traffic shaping over general-purpose, non-deterministic networks without specialized hardware. V-TSN runs in real time alongside the unmodified application stack, serving both as a development-time emulation tool and as a cost-efficient deployment option where relaxed timing is acceptable. In a cloud-based deployment, V-TSN achieves an average clock offset below 200 microseconds, it isolates time-critical traffic through a virtual Time-Aware Shaper (TAS), and it enforces per-class bandwidth reservations through a virtual Credit-Based Shaper (CBS).
\end{abstract}


\section{Introduction}
Time-Sensitive Networking (TSN)~\cite{TSN} is a set of IEEE 802.1 standards that extend Ethernet with deterministic communication, enabling time synchronization, bounded latency, low jitter, and high reliability for time-critical distributed systems. It integrates functionalities such as precise time synchronization~\cite{IEEEAS}, Time-Aware shaper (TAS)~\cite{IEEEQbv}, Credit-Based Shaper (CBS)~\cite{IEEEQav}, and frame preemption~\cite{IEEEQbu} to enable predictable data exchange over standard Ethernet. TSN is a key enabling technology across domains such as industrial, in-vehicle, aerospace, and robotic, where distributed components must operate in tight coordination based on a shared notion of time to ensure correctness, safety, and reliability~\cite{IntroTSN}.

Modern distributed systems are increasingly deployed on commodity and virtualized platforms, where applications run as virtual machines or containers on nodes ranging from edge platforms to on-chip systems. These nodes are typically interconnected through general-purpose networks such as the Internet or cloud infrastructures, which provide neither deterministic communication nor native TSN capabilities~\cite{Struhar2020}. In such environments, clocks are at best loosely synchronized through software protocols such as NTP, or not synchronized at all, and there is no synchronized TSN protocol stack, no deterministic forwarding, and no time-aware traffic shaping. Achieving genuine TSN determinism and TSN-grade synchronization on these platforms requires specialized hardware, such as TSN-capable network interfaces and switches with hardware timestamping, which significantly increases system cost and deployment effort~\cite{TSNCost}. This level of precision, however, is not always necessary; in many practical scenarios, moderate timing accuracy is sufficient for correct system behavior~\cite{ULL}.

Beyond cost, faithfully reproducing TSN behavior during development and validation is itself challenging, particularly because in early design stages the target TSN hardware is often not yet available. Thus, validation must be carried out on virtual platforms rather than on the final infrastructure. This is especially evident in domains such as automotive~\cite{TSNIVN} and aerospace~\cite{TSNaero}, where extensive validation is performed in virtual environments before integration into physical systems and where timing behavior is critical for correct and safe operation. Simulation-based tools~\cite{OMNET,INSIM} support offline design and analysis of TSN systems, but they run in simulated time, far slower than real time, and cannot execute the actual application software. They therefore cannot reproduce the runtime behavior of a complete system over real distributed infrastructure, which is what is needed to validate timing interactions and functional correctness prior to deployment.

To address the cost and flexibility limitations of hardware-based network infrastructures, the networking community is increasingly adopting paradigms such as Network Function Virtualization (NFV) and Software-Defined Networking (SDN)~\cite{NFV_SDN_Overview}. NFV enables the implementation of network functions as software-based Virtual Network Functions (VNFs), allowing them to run on commodity computing platforms instead of dedicated hardware appliances. Similarly, SDN decouples the control and data planes, enabling centralized and programmable network management. These approaches improve flexibility, scalability, and resource utilization, while also reducing deployment and operational costs~\cite{ETSI}. A wide range of network functions, including routing~\cite{SDNSurvey}, firewalling~\cite{VirtualFirewall}, load balancing, and traffic monitoring~\cite{SDBalancing}, have been successfully virtualized using these paradigms~\cite{NFVUseCase}. However, only a limited number of studies, reviewed in Section~\ref{sec:related_work}, have investigated the virtualization of TSN functionalities.

\begin{figure}[t]
    \centering
    \includegraphics[width=\columnwidth]{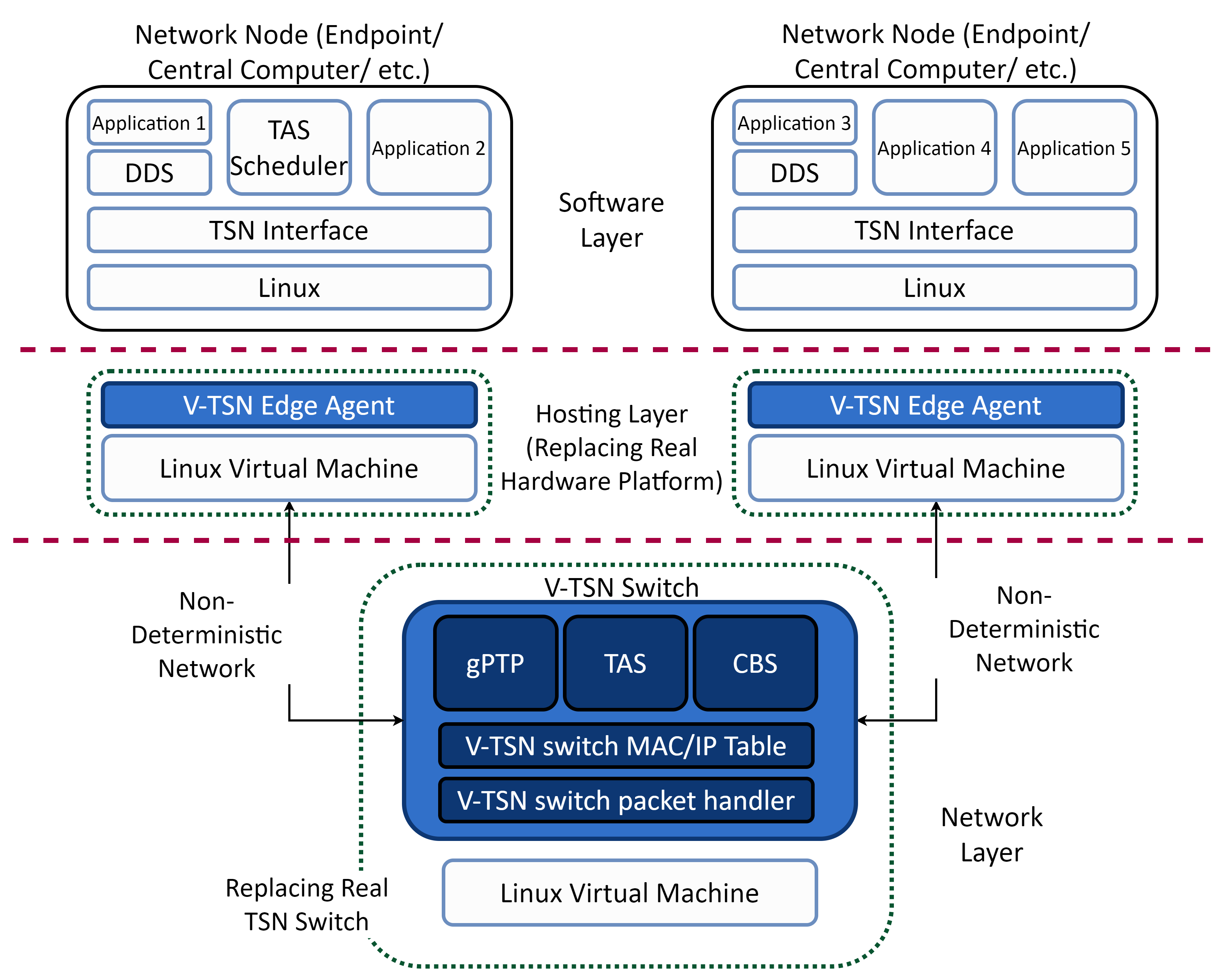}
    \caption{The V-TSN overlay. Each endpoint and the switch run as software components on Linux virtual machines, interconnected over a general-purpose, non-deterministic network. The V-TSN switch replaces a physical TSN switch and implements gPTP synchronization, TAS, and CBS in software.}
    \label{fig:uc1}
\end{figure}

This paper introduces V-TSN, a software-defined overlay that realizes core TSN functionalities, namely gPTP-based synchronization, TAS, and CBS, over general-purpose, non-deterministic networks without requiring specialized hardware. Unlike simulation, V-TSN runs in real time together with the unmodified application stack, and, unlike hardware TSN, it targets relaxed sub-millisecond timing precision rather than nanosecond-level guarantees. V-TSN supports two use cases:
\begin{itemize}[leftmargin=*, nosep]
\item \textbf{UC1, development-time emulation:} V-TSN reproduces TSN behavior so that a system can be validated before integration on physical TSN infrastructure, where we assume that the virtualized system runs at \emph{implementation speed}, so that its timing reflects the eventual hardware deployment.
\item \textbf{UC2, cost-efficient deployment:} Where full TSN hardware is unnecessary and the application tolerates relaxed timing, V-TSN provides TSN-related timing behavior directly over general-purpose networks, for example in distributed sensor data collection, multimedia streaming, and coordination between edge and cloud.
\end{itemize}
Fig.~\ref{fig:uc1} shows the V-TSN overlay, in which endpoints and the switch run as standard computing nodes, such as cloud-based virtual machines, interconnected over a general-purpose network. We demonstrate the feasibility of the proposed approach through a real-world cloud-based deployment, evaluating synchronization accuracy, traffic isolation through virtual TAS, and bandwidth regulation through virtual CBS. The software described in this paper is available via the TU/e ES GitHub repository (https://github.com/TUE-EE-ES).

The remainder of this paper is organized as follows. Section~\ref{sec:related_work} reviews related work. Section~\ref{sec:platform} presents the design of the proposed V-TSN platform. Section~\ref{sec:eval} evaluates the proposed approach. Finally, Section~\ref{sec:conclusions} concludes.

\section{Related Work}\label{sec:related_work}
Realizing TSN behavior over general-purpose networks requires both time synchronization and traffic shaping, since time-aware shaping is meaningful only when nodes are synchronized. Prior work addresses parts of this problem through TSN simulation, hardware testbeds, endpoint synchronization, and, more recently, the virtualization of TSN over commodity platforms, which we review and contrast with V-TSN.

TSN simulation frameworks such as OMNeT++~\cite{OMNET} and INSIM~\cite{INSIM} enable offline evaluation of scheduling strategies and protocol behavior. However, they run in simulated time, far slower than real time, and cannot connect real applications, so they are unsuitable for development workflows that require interaction with actual software.

Hardware testbeds such as~\cite{Cross-validating2024} and EnGINE~\cite{Engine} combine Commercial Off-The-Shelf Ethernet devices with orchestration frameworks for high-fidelity measurements. They depend on physical TSN-capable infrastructure and fixed setups, which we avoid in this work.

Protocols such as Linux PTP~\cite{linuxptp} and PTPd~\cite{ptpd} assume hardware support for timestamping and synchronize network endpoints only. In contrast, V-TSN implements gPTP, which is part of the TSN standard, entirely in software. It synchronizes endpoints as well as intermediate virtualized TSN switches. The latter also implement TAS and CBS traffic shaping.

Closest to our work is a recent research line that virtualizes TSN over commodity platforms. Sasiain et al.~\cite{SasiainNFV} propose to integrate TSN into the ETSI NFV framework, but their proposal is unimplemented, limited to TAS, and assumes TSN hardware. Zhang et al.~\cite{ZhangVTSN} treat virtualized TSN as a control-plane problem of joint admission control and VNF embedding, evaluated in simulation rather than as a running overlay. Most similar, KuberneTSN~\cite{KuberneTSN} provides a deterministic container overlay with a userspace implementation of TAS and kernel-bypass acceleration for Kubernetes environments. However, KuberneTSN virtualizes packet scheduling at the container host and relies on an existing TSN infrastructure for synchronization and switching. Furthermore, it is evaluated on a controlled local testbed with externally synchronized hosts and a physical TSN-compliant switch. Instead, V-TSN virtualizes the complete TSN communication substrate entirely in software. The proposed platform virtualizes both the endpoints and the switch and it operates over non-deterministic general-purpose networks without TSN hardware. Table~\ref{tab:comparison} summarizes the differences between existing approaches and V-TSN.

\begin{table}[t]
\centering
\caption{Comparing TSN virtualization approaches}
\label{tab:comparison}
\footnotesize
\begin{threeparttable}
\renewcommand{\arraystretch}{1.1}
\setlength{\tabcolsep}{5pt}
\begin{tabular}{lcccc}
\hline
\textbf{Feature} &
\makecell{Sasiain\\\cite{SasiainNFV}} &
\makecell{Zhang\\\cite{ZhangVTSN}} &
\makecell{KuberneTSN\\\cite{KuberneTSN}} &
\makecell{V-TSN\\(Ours)} \\
\hline

Prototype implementation &
\xmark & \xmark & \cmark & \cmark \\

gPTP virtualization &
\xmark & \xmark & \xmark & \cmark \\

Virtual TSN switch &
\xmark & \xmark & \xmark & \cmark \\

TAS &
Partial & \xmark & \cmark & \cmark \\

CBS &
\xmark & \xmark & \xmark & \cmark \\

\makecell[l]{No TSN hardware needed} &
\xmark & N/A & \xmark\tnote{1}
& \cmark \\

WAN deployment &
\xmark & \xmark & \xmark & \cmark \\
\hline
\end{tabular}
\begin{tablenotes}
\item[1] KuberneTSN runs on commodity hosts, but its determinism is demonstrated on a testbed with TSN-capable NICs and a physical TSN switch.
\end{tablenotes}
\end{threeparttable}
\vspace*{-7pt}
\end{table}

\section{V-TSN Platform Design}\label{sec:platform}
In V-TSN, both the endpoints and the switches are software components that can run on general-purpose hosts and be placed anywhere in the network. 
A virtual switch is just another software component, a virtual machine (VM) that may even run on an endpoint. To the underlying network, it carries only ordinary traffic that standard network switches forward, like any other switch. Lightweight software agents run at the endpoints, while the core networking logic resides in the virtual V-TSN switch, on top of which the TSN functionalities are layered.

The following subsections describe, first, the virtual packet switching that carries all traffic, then the software realization of gPTP for synchronization, and finally the TAS and CBS traffic shaping implemented within the V-TSN switch.

\subsection{Virtual Packet Switching Mechanism}
\label{subsec:switching}
The basis of the V-TSN platform is a virtual packet switching mechanism that enables communication between distributed endpoints over a general-purpose network. Since the underlying network does not provide native TSN support, the proposed approach introduces an overlay communication layer in which Ethernet frames are encapsulated, transmitted, and processed through software-defined components at the endpoints and the virtual V-TSN switches.

\begin{figure*}[t]
    \centering
    \includegraphics[width=0.9\linewidth]{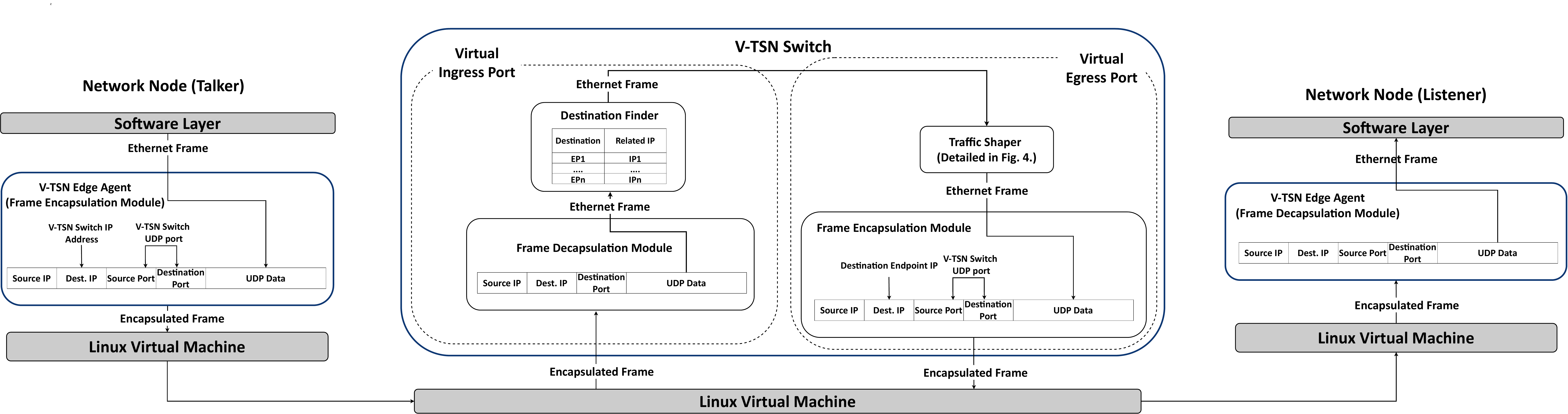}
    \caption{Virtual packet switching mechanism in the V-TSN platform.}
    \label{fig:switching}
\end{figure*}

Fig.~\ref{fig:switching} illustrates this packet processing pipeline. At the talker, a frame encapsulation module wraps each outgoing Ethernet frame in a UDP/IP packet addressed to the V-TSN switch, enabling transport over standard IP networks. At the switch, (i) incoming packets are decapsulated, (ii) a virtual destination resolution module selects the output endpoint from a mapping table of logical endpoints to IP addresses and ports, (iii) the frame is optionally processed by the traffic shaping mechanisms, and (iv) it is then re-encapsulated and forwarded. At the listener, a decapsulation module recovers the original Ethernet frame and an application mapping mechanism delivers it to the correct endpoint component using header fields such as ports. This effectively emulates a layer-2 switch over a layer-3 network while integrating TSN traffic control.

\subsection{Virtual gPTP Synchronization}
Time synchronization in TSN is provided by gPTP, which aligns the local clocks of distributed nodes with a grandmaster reference clock. Unlike end-to-end protocols, gPTP is a hop-to-hop protocol: it measures link delay on each hop through a peer-delay exchange (\emph{Pdelay\_Req}/\emph{Pdelay\_Resp}) and distributes the reference time through \emph{Sync} and \emph{Follow\_Up} messages, which yields higher synchronization accuracy.

In the V-TSN platform, gPTP is implemented by handlers that run as ordinary applications on the endpoints and the V-TSN switch. One Linux VM acts as the grandmaster node. Because gPTP uses normal Ethernet packets, the exchanges are encapsulated and transported over the virtualized TSN network as described above like regular data traffic, avoiding a separate synchronization channel and preserving the structure of gPTP communication over a non-deterministic network.
 
\begin{figure}[t]
    \centering
    \includegraphics[width=\linewidth]{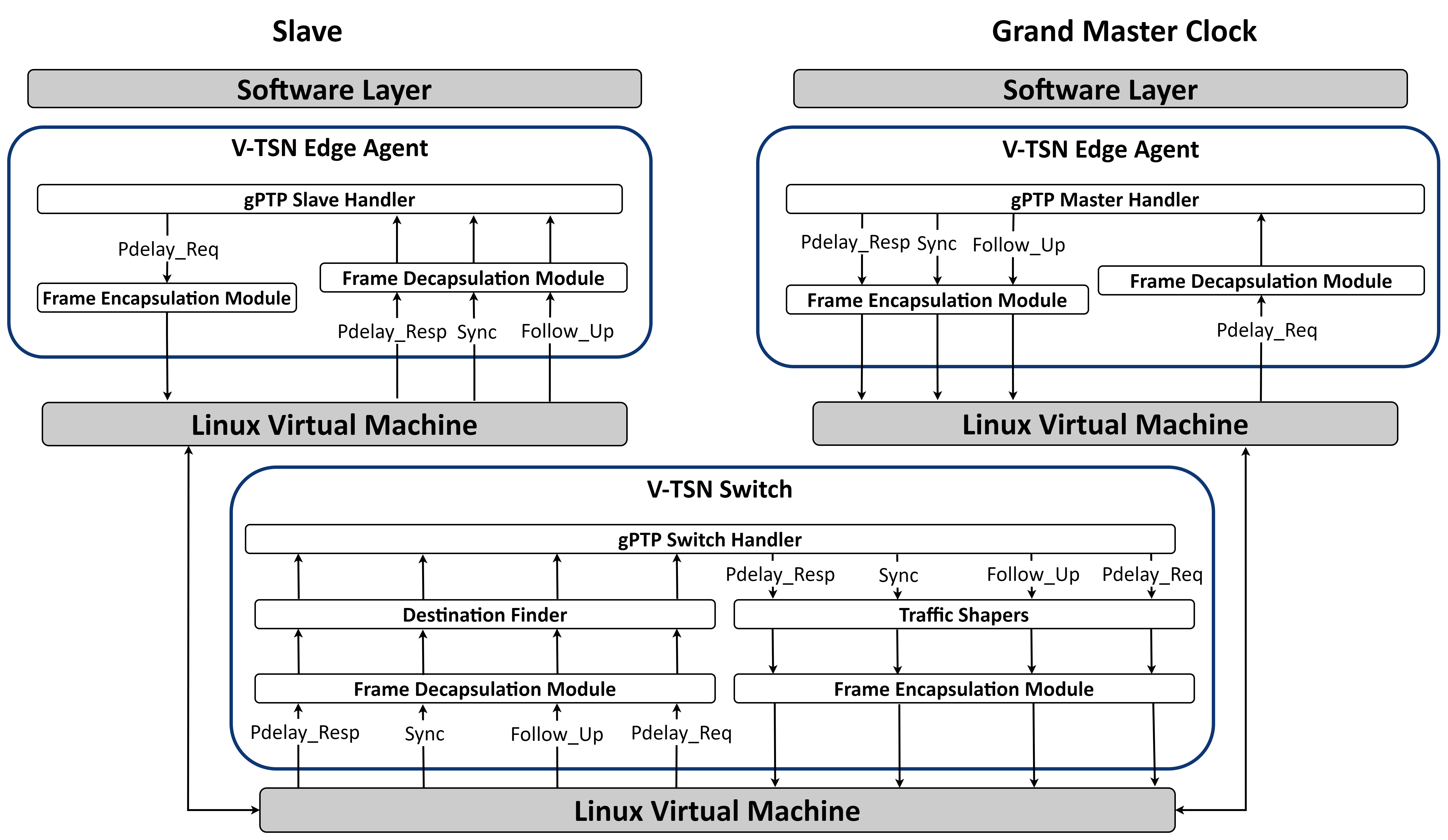}
    \caption{Virtual gPTP synchronization in the V-TSN platform.}
    \label{fig:gptp}
\end{figure}

\subsection{Virtual Traffic Shaping}
Traffic shaping in TSN controls the timing and rate of packet transmissions to achieve predictable communication behavior. In the V-TSN platform, traffic shaping is implemented within the V-TSN switch, providing a software-based realization of TSN scheduling over a non-deterministic network.

\begin{figure}[t]
    \centering
    \includegraphics[width=0.75\linewidth]{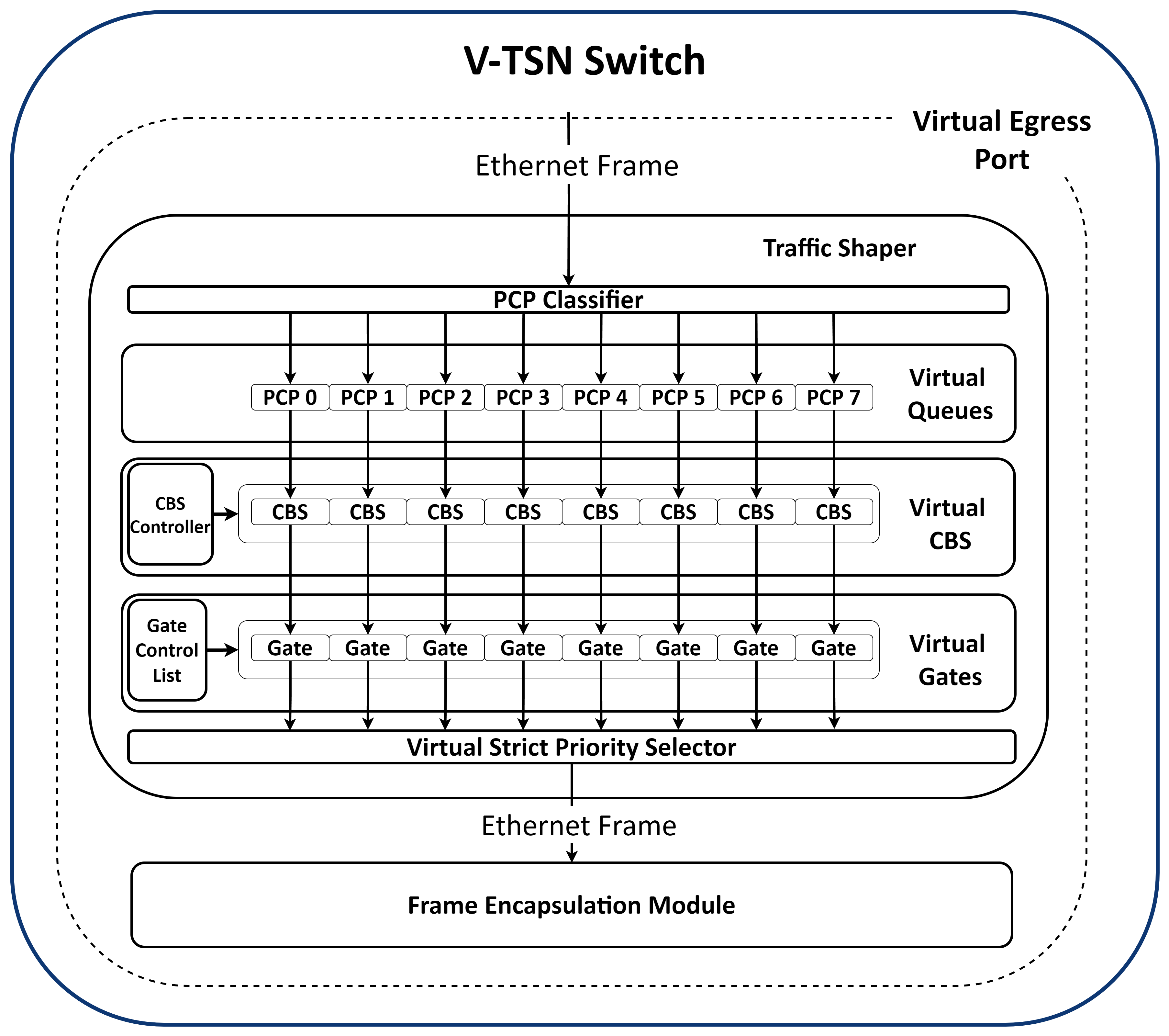}
    \caption{Virtual traffic shaping architecture inside the V-TSN switch, detailing the traffic shaper of Fig.~\ref{fig:switching}.}
    \label{fig:shaping}
\end{figure}

Fig.~\ref{fig:shaping} illustrates the architecture of the virtual traffic shaping module. At the egress of the shaping pipeline, Ethernet frames are first classified by their Priority Code Point (PCP) values and mapped to one of eight virtual queues corresponding to the standard priority levels (PCP 0 to PCP 7),  representing the logical separation of traffic classes in TSN.

Two shaping mechanisms are applied within the switch: CBS and TAS. CBS is implemented through a set of virtual CBS modules. A CBS controller regulates the transmission rate of each queue by maintaining and updating credit values, ensuring bandwidth sharing and smooth traffic behavior. After CBS, the TAS mechanism is realized using a virtual Gate Control List (GCL), where each queue is associated with a gate that opens or closes according to a predefined schedule, enabling time-based control so that high-priority traffic is transmitted within specific time windows. The GCL gates open and close at specific time instants based on the global time base provided by gPTP.
The accuracy of this alignment is limited by the gPTP synchronization error and each switch hop can add up to one GCL cycle of gating delay, both of which contribute to the end-to-end latency of time-critical traffic.

After passing through the CBS and TAS stages, packets are forwarded to a virtual strict-priority selector, which determines the next packet to transmit based on priority. The selected packet is then transmitted through the underlying network using the encapsulation mechanism described earlier.

Although these mechanisms are implemented in software and do not provide the same level of determinism and accuracy as hardware-based TSN, they preserve the logical behavior of TSN traffic shaping. This allows the V-TSN platform to emulate scheduling effects and traffic interactions, enabling meaningful evaluation of time-sensitive applications.

\section{Evaluation and Results}\label{sec:eval}
We evaluate the V-TSN platform deployed on a commercial cloud provider over the public Internet, with two objectives. First, we measure the clock synchronization accuracy achieved by the virtual gPTP mechanism, which determines the quality of the shared time base available to applications. It is  relevant to both use cases highlighted in the introduction. Second, for the emulation use case, UC1, we validate that the virtual TAS and CBS reproduce the behavior expected of TSN traffic shapers. The aim of this second part is to confirm that the platform exhibits the characteristic TSN shaping behavior.

\subsection{Experimental Setup}\label{subsec:exp_setup}
To evaluate V-TSN under realistic conditions, we deploy it on commercial cloud infrastructure rather than a controlled testbed. As summarized in Table~\ref{tab:exp_setup}, the experiments use four cloud VMs acting as a grandmaster clock (and also as endpoint EP0), the V-TSN switch, and two endpoints EP1 and EP2. Three nodes are co-located in the same region, while EP2 is in a different country, introducing wide-area variability; the co-located nodes therefore represent a favorable setup and EP2 dominates the wide-area behavior. All machines communicate over UDP across the public Internet, with no TSN-capable hardware and no controlled network conditions. So the overlay runs over a genuinely non-deterministic wide-area network. All endpoint traffic is routed through the V-TSN switch, representative of a real deployment in which distributed nodes are interconnected through general-purpose cloud networks.

\begin{table}[t]
\centering
\caption{Experimental setup summary}
\label{tab:exp_setup}
\footnotesize
\setlength{\tabcolsep}{4pt}
\begin{tabular}{|c|c|c|c|}
\hline
VM & Location & Configuration & Role \\ \hline
VM1 & Germany & Hetzner, Linux, 2 vCPU, 4 GB & Grandmaster (EP0) \\
VM2 & Germany & Hetzner, Linux, 2 vCPU, 4 GB & V-TSN Switch \\
VM3 & Germany & Hetzner, Linux, 2 vCPU, 4 GB & Endpoint 1 (EP1) \\
VM4 & Finland & Hetzner, Linux, 2 vCPU, 4 GB & Endpoint 2 (EP2) \\ \hline
\end{tabular}
\end{table}

\subsection{Time Synchronization Accuracy}
\label{subsec:sync}
\begin{figure}[t]
    \centering
    \begin{subfigure}[t]{0.44\columnwidth}
        \centering
        \includegraphics[width=\linewidth]{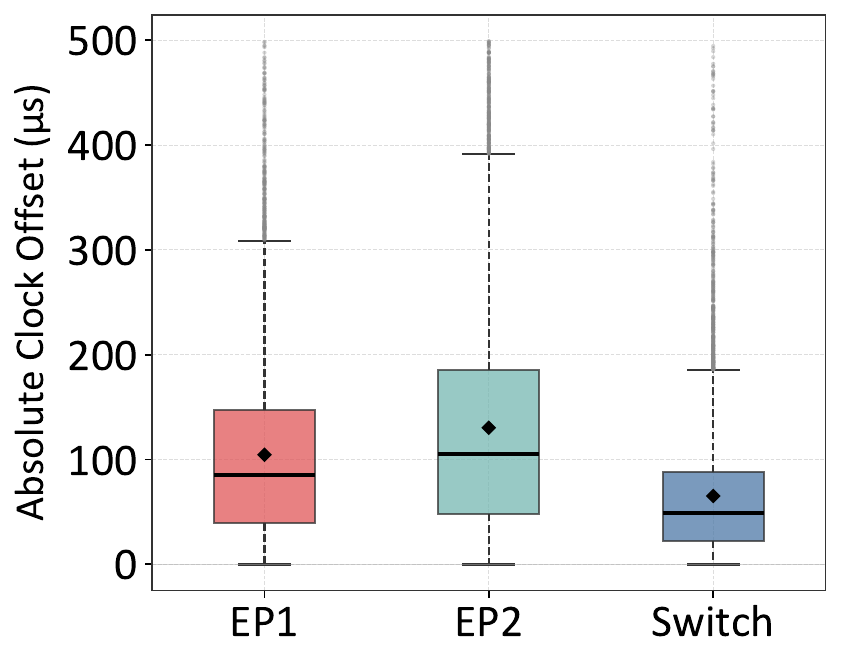}
        \caption{Offset within 500~$\mu$s.}
        \label{fig:clockdrift_boxplot}
    \end{subfigure}
    \hfill
    \begin{subfigure}[t]{0.49\columnwidth}
        \centering
        \includegraphics[width=\linewidth]{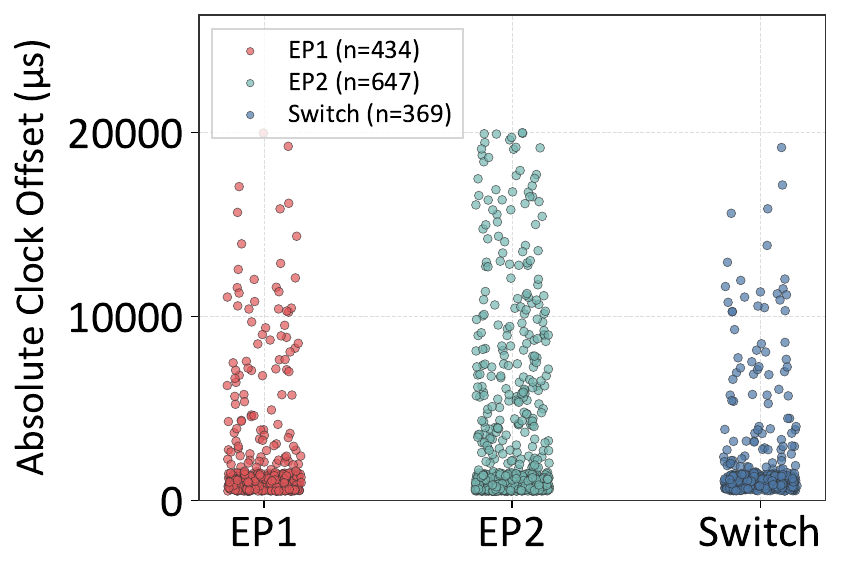}
        \caption{Outliers above 500~$\mu$s.}
        \label{fig:clockdrift_dotplot_outliers}
    \end{subfigure}
    \caption{Clock offset across nodes over approximately 10,000 samples per node: (a) the main distribution and (b) the outliers exceeding 500~$\mu$s.}
    \label{fig:clockdrift}
\end{figure}

We evaluate synchronization accuracy through the clock offset of each node relative to the grandmaster. At each synchronization instant, when a node receives a new reference time from the grandmaster, it computes its offset as the received grandmaster time minus its own local time at that moment. We report the absolute value of this offset. Fig.~\ref{fig:clockdrift_boxplot} shows a boxplot of 10,000 such offset samples per node collected over 80 hours, with the diamond marking the mean. For readability, samples exceeding 500~$\mu$s are excluded here and shown separately in Fig.~\ref{fig:clockdrift_dotplot_outliers}.

As shown in Fig.~\ref{fig:clockdrift_boxplot}, the average clock offset remains below 200~$\mu$s for all nodes, with the V-TSN switch the most tightly synchronized, consistent with its co-location with the grandmaster. EP2, which is deployed in a different geographical location, exhibits higher offset values and larger variation compared to the other nodes. Increased geographical distance and the associated network variability have a direct impact on synchronization accuracy.

Fig.~\ref{fig:clockdrift_dotplot_outliers} presents the offset samples greater than 500~$\mu$s. All three nodes exhibit such deviations, with EP1, EP2, and the switch showing 434, 647, and 369 samples above this threshold. These outliers are consistent with transient congestion events on the public Internet, suggesting that WAN latency variability is the dominant factor in worst-case synchronization error. Such deviations represent no more than about 6.5\% of each node's samples and the system recovers to normal synchronization accuracy once network conditions stabilize.

Overall, V-TSN achieves sub-millisecond synchronization under typical conditions, with occasional larger deviations caused by the non-deterministic network.

\subsection{Analysis of Virtual TAS Behavior}
\label{subsec:tas}

In this section, we evaluate the behavior of the virtual TAS in terms of traffic isolation and latency. To this end, two traffic streams representing different traffic classes are generated by separate senders and forwarded by the V-TSN switch to a common receiver. Endpoint 1 (EP1) sends a best-effort stream (Traffic Class 0, TC0) and the grandmaster (EP0) sends a time-critical stream (Traffic Class 3, TC3), both arriving independently at the switch, which forwards them to EP2, where the one-way latency of each packet is recorded.

The experiment is designed to create contention at the egress of the V-TSN switch. A bursty Best-Effort (TC0) traffic pattern is introduced to generate congestion on the shared virtual link, potentially blocking Time-Critical (TC3) packets. TC0 traffic is generated as 20-packet bursts every 300~ms, while TC3 traffic follows a steady transmission pattern of 50 packets per second. The egress virtual port is configured with a transmission rate of 1.5~Mb/s. The payload sizes are set to 1200~B for TC0 and 64~B for TC3.

Two configurations are considered. In the first configuration (without TAS), all gates remain open throughout the cycle, resulting in a shared FIFO transmission behavior where TC3 packets must wait behind TC0 traffic. In the second configuration (with TAS), a periodic schedule with a cycle of 14~ms is applied. Within each cycle, TC3 is assigned a dedicated transmission window of 7~ms, while TC0 is assigned the remaining 7~ms. This ensures that TC3 traffic obtains exclusive access to the link during its allocated time window.

\begin{figure}[t]
    \centering
    \hspace*{5mm}\begin{subfigure}[t]{0.4\columnwidth}
        \centering
        \includegraphics[width=\linewidth]{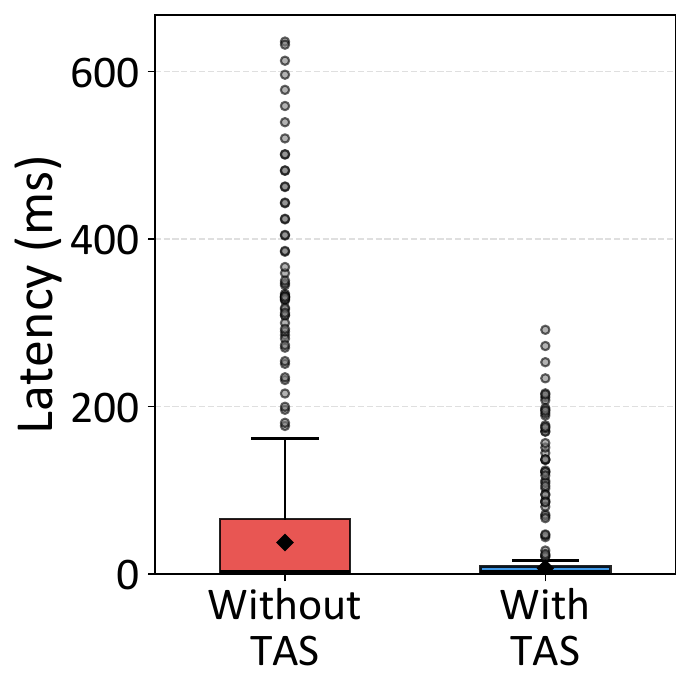}
        \caption{Time-Critical latency with and without TAS.}
        \label{fig:tas_boxplot_tc3}
    \end{subfigure}
    \hfill
    \begin{subfigure}[t]{0.4\columnwidth}
        \centering
        \includegraphics[width=\linewidth]{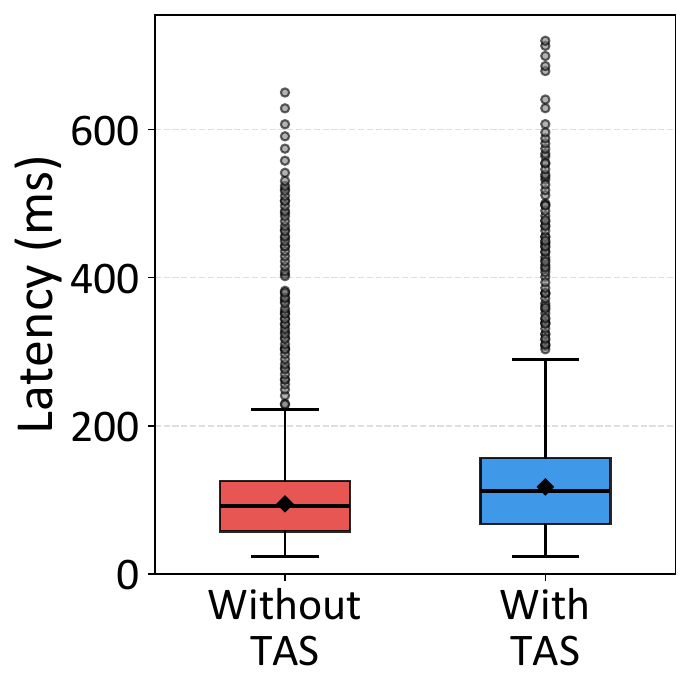}
        \caption{Best-Effort latency with and without TAS.}
        \label{fig:tas_boxplot_tc0}
    \end{subfigure}\hspace*{5mm}
    \caption{Impact of TAS on latency across traffic classes.}
    \label{fig:tas_boxplots}
\end{figure}

The resulting latency distributions are shown in Fig.~\ref{fig:tas_boxplots}. Fig.~\ref{fig:tas_boxplot_tc3} shows that TAS sharply reduces the mean and worst-case latency of the time-critical TC3 stream and tightens its distribution. Without TAS, TC3 packets are frequently blocked behind bursty TC0 traffic at the switch egress, giving a mean latency of about 38~ms and a worst case above 600~ms. With TAS, TC3 is served within its dedicated window and shielded from TC0 interference, reducing the mean to about 7~ms and the worst case to about 292~ms. The residual worst case reflects the wide-area path to EP2 rather than switch contention.

Fig.~\ref{fig:tas_boxplot_tc0} shows that best-effort TC0 latency increases when TAS is enabled, with the mean rising from about 95~ms to 118~ms and the worst case from about 650~ms to 720~ms, as best-effort traffic is held back while the TC3 gate is open.

Overall, the virtual TAS reproduces the key behavior of hardware TAS, providing temporal isolation and prioritized access for time-critical traffic over a non-deterministic network. Note that the underlying Internet adds delay and variability that must be accounted for when designing the GCL schedule.

\subsection{Analysis of Virtual CBS Behavior}
\label{subsec:cbs}

In this part, we evaluate the behavior of the virtual CBS and its ability to enforce bandwidth reservations across traffic classes. Similar to the TAS experiment, two traffic streams corresponding to different traffic classes are generated. In this case, two independent senders are used, where the grandmaster node (EP0) is additionally utilized as a traffic source to generate one of the streams.

The egress virtual port is configured with a transmission rate of 10~Mb/s. CBS parameters are configured such that TC0 (best effort) is assigned an idleSlope corresponding to 7~Mb/s, while TC3 (high priority) is assigned 2~Mb/s, leaving remaining bandwidth as headroom. Both traffic streams are generated with time-varying sending rates to create dynamic load conditions that periodically exceed the reserved bandwidth, allowing the shaping behavior to be clearly observed.

\begin{figure}[t]
    \centering
    \includegraphics[width=0.9\columnwidth]{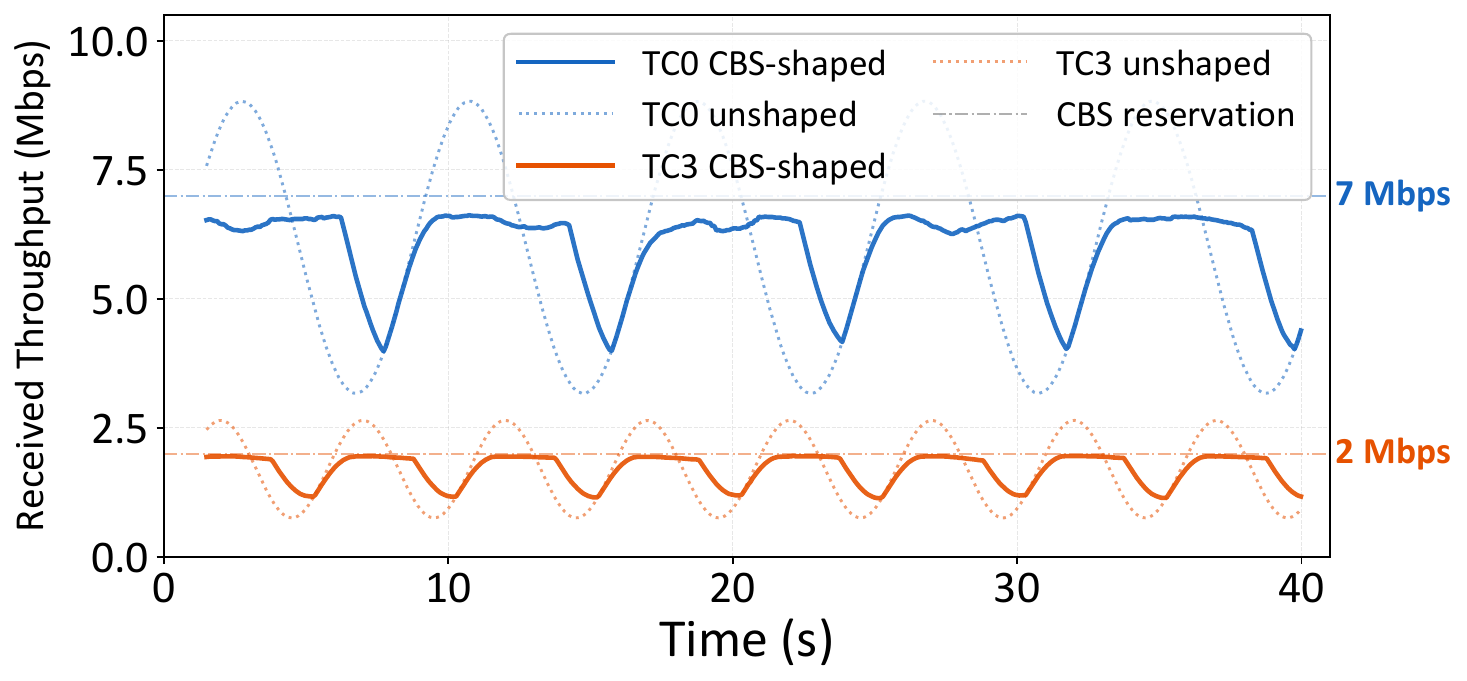}
    \caption{Effect of CBS on throughput for different traffic classes.}
    \label{fig:cbs_result}
\end{figure}

Fig.~\ref{fig:cbs_result} shows the throughput over time. The unshaped traffic follows the input rate and exceeds the desired bandwidth limits during peaks. In contrast, the CBS-shaped traffic is effectively constrained by the configured reservations. In particular, when the input rate exceeds the assigned bandwidth, the throughput is clipped at the configured limit, while for lower input rates, the traffic passes through without modification.

The shaping occurs for both traffic classes simultaneously, confirming that the virtual CBS enforces per-class bandwidth guarantees despite a shared output port, reproducing the rate limiting and fair allocation of hardware CBS under contention. Network-induced delay has less impact on CBS than on TAS, since CBS regulates input rate rather than relying on strict time slots, though software processing overhead may cause slight deviations from ideal CBS.

\section{Conclusion and Future work}\label{sec:conclusions}
In this paper, we introduced V-TSN, a software-defined overlay that realizes core TSN functionalities over general-purpose, non-deterministic networks without specialized hardware. The same platform serves two use cases: as a development-time emulation tool for validating TSN-based systems before the target hardware is available, and as a cost-efficient deployment option for applications that tolerate relaxed timing over general-purpose or cloud networks.

A real-world cloud deployment confirms that V-TSN reproduces the essential behavior of hardware TSN mechanisms in software: the virtual gPTP keeps the average clock offset below 200~$\mu$s, the virtual TAS isolates time-critical traffic and reduces its mean and worst-case latency under contention, and the virtual CBS enforces per-class bandwidth reservations on a shared output port.

As future work, we plan to combine V-TSN with platforms that virtualize the computation side of distributed systems, such as virtual ECU and application environments. Pairing a virtualized TSN network with virtualized compute would allow an entire time-sensitive system, both its applications and its communication, to be developed and validated end-to-end before any physical hardware is available. An example platform is the AUMOVIO~\cite{aumovio} in-vehicle virtualization platform, for which V-TSN could provide the networking layer in a fully virtual end-to-end validation flow for in-vehicle applications.

\section*{Acknowledgment}

This work has received funding from the European Chips Joint Undertaking under Framework Partnership Agreement No 101139789 (HAL4SDV).

\end{document}